# Dynamic Kerr Effects and Spectral Weight Transfer in the Manganites

S.A. McGill[*], R.I. Miller[*], O.N. Torrens[*], A. Mamchik[†], I-Wei Chen[†] and J.M. Kikkawa[*]

[*]*Department of Physics & Astronomy, University of Pennsylvania, Philadelphia, PA 19104*

[†]*Department of Materials Science & Engineering, University of Pennsylvania, Philadelphia, PA 19104*

Abstract

We perform pump-probe Kerr spectroscopy in the colossally magnetoresistive manganite $Pr_{0.67}Ca_{0.33}MnO_3$. Kerr effects uncover surface magnetic dynamics undetected by established methods based on reflectivity and optical spectral weight transfer. Our findings indicate the connection between spin and charge dynamics in the manganites may be weaker than previously thought. Additionally, important differences between this system and conventional ferromagnetic metals manifest as long-lived, magneto-optical coupling transients, which may be generic to all manganites.

The colossally magnetoresistive (CMR) manganites have attracted much attention due to their unconventional insulator-metal (IM) transitions, where competing tendencies such as double exchange spin alignment, Jahn-Teller polaron formation, and charge ordering give rise to sensitive phase instabilities [1]. The seemingly intrinsic phase-separation of these highly correlated systems presents challenges for their characterization [2], so that spectroscopies capable of identifying contributions from electronic and magnetic heterogeneity are of great interest. Time-resolved optical probes are capable of exploring couplings between spin, charge and lattice degrees of freedom [3], and the central role of spin in CMR has motivated several investigations where transient reflectivity and/or transmissivity are interpreted as *quantitative* measurements of manganite spin dynamics [4,5]. In this Letter, time-resolved Kerr effects suggest a less rigid connection between spin and charge dynamics and reveal heterogeneous magnetic phases not detected in simultaneous reflectivity measurements. Additional measurements further reveal that magneto-optical (M-O) coupling transients, extensively studied in conventional ferromagnets for their importance on picosecond time scales [6-9], can last three orders of magnitude longer in the manganites due to their unique charge dynamics.

Our studies contrast with earlier approaches based on time-resolved reflectivity and transmissivity, where spectral weight redistribution between the mid- and far-IR is assumed to mirror ferromagnetic demagnetization on sub-nanosecond timescales. This latter magnetic interpretation of 'dynamic spectral weight transfer' (DSWT) appears motivated by correlations between spectral weight transfer and magnetization observed under quasi-equilibrium conditions [10]. Yet the dynamical connection between charge and spin in the manganites is of fundamental interest in itself, and it is worthwhile to test these notions by more traditional measures of non-equilibrium magnetization such as the Kerr effect. We show that Kerr spectroscopy reveals surface spin dynamics not seen in transient reflectivity, underscoring the importance of using both rotation and ellipticity to distinguish Kerr effects unrelated to magnetization dynamics [11].

Our samples are bulk $Pr_{0.67}Ca_{0.33}MnO_3$ (PCMO) polycrystalline ceramics prepared by the standard sintering route and mechanically polished using diamond paste. Supplemental data are also taken on $La_{0.7}Ca_{0.3}MnO_3$ (LCMO) prepared by the same methods. For time-resolved experiments, we employ sub-picosecond 1 kHz pump and probe pulses at 3.1 eV (260 $\mu$J cm$^{-2}$ per pulse, except where noted) and 1.55 eV [12], respectively. The linearly cross-polarized pump and probe beams focus to a 1.5 mm diameter spot in a 10° non-collinear, near-normal incidence arrangement with B along the sample normal. Probe Kerr rotation employs a broadband polarization rotator followed by a polarization beam splitter, whereas ellipticity requires an additional quarter wave plate. Motorized rotation of the optical elements compensates for the field-dependent static rotation from the cryostat windows. A low-noise photodiode bridge detects the polarization-split light, and the diode difference (sum) at the pump's chopper frequency of 97 Hz measures pump-induced rotation or ellipticity (reflectivity). Physical units are obtained through point-by-point normalization of these quantities by the reflected probe intensity.

Kerr rotation, $\theta$, and ellipticity, $\eta$, are linearly related to the magnetization, M, by M-O coupling constants, $f$ and $g$, respectively. Consequently, transient changes in the Kerr signals reflect pump-induced changes in the magnetization, as well as in $f$ and $g$,

$$\Delta\theta = f \cdot \Delta M + \Delta f \cdot M + C_1 \qquad (1)$$

$$\Delta\eta = g \cdot \Delta M + \Delta g \cdot M + C_2 \tag{2}$$

where $C_1$ and $C_2$ account for non-magnetic and higher order terms. Here, we introduce new measurement procedures that account for resistive and magnetic hysteresis typical of CMR manganites (Fig. 1). Note that the first and second terms in Eqns. (1-2) remain after measurements at opposite magnetic fields (B,−B) are subtracted, assuming these fields are set in a manner that prepares magnetization-reversed but otherwise identical states. To ensure the latter, data is first taken at B after a field-increasing sweep from −7 T to B at 0.5 T/min, and field-reversed data is then obtained *in the same resistive state* after a decreasing sweep from +7 T to −B at −0.5 T/min. For rotation, we denote these measurements $\Delta\theta^{inc}(B)$ and $\Delta\theta^{dec}(-B)$, respectively, and always report values of $\Delta\tilde{\theta}(B) \equiv [\Delta\theta^{inc}(B) - \Delta\theta^{dec}(-B)]/2$. By our convention, B can be positive or negative, and the resistive state corresponding to $\Delta\tilde{\theta}(B)$ is that obtained along an increasing field sweep from −7 T to B (see field pairs in Fig. 1). The same method is used to obtain $\Delta\tilde{\eta}(B)$. Importantly, when M-O coupling dynamics (second terms) are negligible, $\Delta\tilde{\eta}$ and $\Delta\tilde{\theta}$ quantify $\Delta M$ and can be scaled onto each other as $\Delta\tilde{\eta} \cong (g/f)\Delta\tilde{\theta}$ [6].

These Kerr measurements uncover transient magnetic behaviors and phase heterogeneity unanticipated by bulk measurements such as resistivity or magnetization, and undetected by transient reflectivity ($\Delta R/R$). Figure 2a gives evidence for pump-induced demagnetization (PID) of ferromagnetism not only in the low-resistance, ferromagnetic state, but also in the insulating state, where bulk PCMO is antiferromagnetic [13]. According to our notation, the low-resistance state at T = 35 K is studied by $\Delta\tilde{\theta}(-3\ T) = [\Delta\theta^{inc}(-3T) - \Delta\theta^{dec}(+3T)]/2$, while the insulating state corresponds to $\Delta\tilde{\theta}(+3\text{ T}) = [\Delta\theta^{inc}(+3\text{T}) - \Delta\theta^{dec}(-3\text{T})]/2$ (see field pairs in Fig. 1). Assuming $\Delta\tilde{\theta}(-3\text{ T}) > 0$ represents pump-induced demagnetization (PID) of the low-resistance, ferromagnetic state ($\Delta M/M < 0$), then the observed $\Delta\tilde{\theta}(+3\text{ T}) < 0$ suggests a relatively large PID in the insulating state as well ($\Delta M/M < 0$ is implied after taking into account a sign reversal in M due to B = -3 T → +3 T). The presence of a minority ferromagnetic surface phase is clearly revealed by varying the magnetic field (Fig. 2b). Here, continuous field scans of $\Delta\tilde{\theta}(B)$ are taken at a fixed time delay ($\Delta t$ = 1500 ps) chosen so that $\Delta\tilde{\theta}(B)$ tracks PID and has negligible contributions from changes in magnetic relaxation times. Naively, one expects PID to scale with the bulk moment (also shown in Fig. 2b), but additional PID contributions at low fields appear for all data below T = 130 K and are sensitive to polishing, indicating a secondary surface phase to which the Kerr effect is particularly responsive. The latter ferromagnetic phase is responsible for PID in the insulating state.

This secondary phase is not detected by bulk magnetization or simultaneous resistivity (Fig. 1). Nor is its presence evident in $\Delta R/R$, even though magnetic interpretations of DSWT would infer $\Delta M$ from $\Delta R/R$ [12]. DSWT predicts a sign change at early time delays across the IM transition, and field scans at $\Delta t$ = 100 ps show that $\Delta R/R$ responds strongly to the *bulk phase* IM transition (Fig. 2b). Yet $\Delta R/R$ does not record large changes in secondary phase demagnetization near zero field (Fig. 2b). From these data it is clear that magnetic interpretations of DSWT can lead to a misinterpretation of the surface physics.

Comparisons in the time domain illustrate further discrepancies between ΔM and ΔR/R. For times greater than ~ 1 ns, $\Delta\tilde{\eta}$ scales onto $\Delta\tilde{\theta}$ and the transient Kerr effects present a quantitative measure of ΔM, as described above. Figures 3a-b show a marked disagreement between magnetic dynamics and that of ΔR/R that is particularly apparent in the insulating state where PID of the secondary ferromagnetic phase occurs. Such contrasting behaviors of ΔM and ΔR/R cannot be explained by variations in sample positioning because ΔR/R is collected simultaneously with the Kerr effects and hence measures exactly the same sample volume. Magnetic interpretations of DSWT in the ferromagnetic state have assumed that picosecond PID processes were irrelevant due to an inefficient spin-lattice coupling, estimating the spin disordering time $\tau_m$ from the rise of the ΔR/R signal toward its first maximum *after the picosecond transient near zero delay* [5]. Following this interpretation, one would conclude that $\tau_m$ in the insulating state is ~1 ns (solid triangles in Fig. 3c) and that PID continues to increase out to 3 ns. Instead, the Kerr data indicate PID is complete much sooner, with ΔM returning to zero for times greater than ~1 ns. Although M-O coupling dynamics Δ*f* and Δg prevent a quantitative interpretation of the Kerr data on sub-nanosecond timescales, expanded in Fig. 3a and 3b insets, both $\Delta\tilde{\theta}$ and $\Delta\tilde{\eta}$ rise an order of magnitude faster than ΔR/R (Fig. 3c) and place a lower limit on $\tau_M$ of ~1-10 ps.

Within the heat-bath model [5], we can make an order-of-magnitude estimate of $\tau_m$ using the magnetic heat capacity ($C_m$) and ferromagnetic spin stiffness (D) in PCMO (x = 0.3) [14]. With $C_m \approx (0.113)k_B(k_BT/D)^{3/2}$, the lattice heat capacity in PCMO (x = ~0.4) ($C_L \approx 14$ J/K-mole) [15], and the value of the spin-lattice coupling energy for manganites (G ≈ $10^{11}$ W/K-mole) [5], we find $\tau_M = C_m C_L/(C_L+C_m)G$ is predicted to be of order 2-10 ps (Fig. 3c). Nonetheless, further studies are necessary to determine the relevance of a variety of other PID mechanisms that can also lead to short $\tau_m$. These include a direct spin-electron thermal channel [7], instantaneous demagnetization by spin-orbit coupling [16], Stoner excitations [18], and minority spin excitations into the d-states which reduce their net magnetization [9], all of which have been discussed in the context of conventional ferromagnetic metals where sub-ps [16-19] or ps [7,8] $\tau_m$ are observed.

We find, however, that spectral changes arising from DSWT produce Kerr effects in the manganites that are fundamentally different from those observed in conventional ferromagnetic metals. The non-scalability of $\Delta\tilde{\theta}$ and $\Delta\tilde{\eta}$ for times less than 1 ns (Figs. 3a-b) indicates significant M-O coupling dynamics (second terms in Eqns. 1-2) on a timescale that is nearly three orders of magnitude longer than observed in conventional ferromagnets such as Ni [8] and $CoPt_3$ [9]. In CMR manganites such as PCMO, DSWT describes pump-induced, broadband spectral weight redistribution associated with the IM transition [20], which can be expected to modify the M-O couplings *f* and *g* on a time scale that can involve lattice structural changes [21]. Such changes would spectrally shift transient reflectance features much more than in conventional ferromagnets, and by varying the probe photon energy ($E_p$) to obtain the $\Delta R/R$ spectrum [22], we find the spectrum indeed changes shape with probe delay (Fig. 4a inset). To quantify the timescale of this change, we obtain the positive-definite area, Σ, between spectra recorded at Δt and a reference scan taken at 4000 ps, first rescaling all spectra to the same average value and then integrating the absolute value of their difference with the reference.

Figure 4a shows that changes in Σ are mirrored by relative changes in $f$ and $g$ (the latter are quantified by $\alpha \equiv \Delta\tilde{\eta} - (g/f)\Delta\tilde{\theta} = (\Delta g - (g/f)\Delta f)M$, the discrepancy between scaled plots of $\Delta\tilde{\theta}$ and $\Delta\tilde{\eta}$ in the main panel of Fig. 3b). This correspondence is seen in both insulating and low-resistance states, and provides strong evidence that dynamical band shifting underlies the observed M-O coupling changes.

This point is corroborated by additional measurements relating changes in equilibrium M-O properties to the manganites' IM transition. Figure 4c shows DC polar Kerr rotation and ellipticity at T = 35 K and 1.55 eV. A comparison of $\theta$ and $\eta$ shows that the equilibrium ratio $f/g$ changes magnitude and sign at the IM transition (dashed vertical lines). These findings suggest that any pump-induced phase excursions relative to the IM boundary can produce time-varying M-O couplings that can obscure magnetization dynamics. DC Kerr measurements also reveal field-dependent changes in $f/g$ in LCMO at T = 270 K (not shown), and the corresponding time-resolved data (shown in Fig. 4b) indicate that rotation and ellipticity are again not proportional on timescales of several nanoseconds.

These changes in $f$ and $g$ are probably generic characteristics of the manganite family where electronic bands and optical transitions shift at the IM boundary. Hence, M-O coupling dynamics can be expected to contribute to Kerr effects in all pump-probe studies of the CMR IM transition and should be carefully included for a proper interpretation of Kerr dynamics [11]. In some cases it may be possible to relate the Kerr and reflectivity transients to changes in lattice temperature or magnetic field. However, in our study the spectral changes appear to be more complex and cannot be explained by heating alone. In the Fig. 4a inset, DC reflectivity changes with respect to T = 35 K are contrasted with transient reflectivity under identical pump conditions. No DC spectral shift can be scaled onto the transient reflectivity at any temperature up to T = 150 K (two representative temperatures are shown) and for any delay from 40 ps to 4 ns.

In conclusion, we find that the Kerr effect is a more accurate probe of surface magnetic dynamics in the manganites than the popular technique based on spectral weight transfer. In the case of PCMO, a secondary magnetic surface phase is probed by the Kerr effect that is not seen in reflectivity. These findings are relevant to transient optical studies on thin films [4] and single crystals [11], where intrinsic and/or surface related phase heterogeneity is commonplace and perhaps unavoidable. Kerr effects unrelated to magnetization dynamics are found to last dramatically longer than in conventional ferromagnets. This phenomenon is linked to dynamic spectral weight transfer, and is likely generic to the CMR manganites and perhaps all correlated electronic systems in which carrier density is coupled to magnetism [23]. This work is supported by NSF-DMR-0094156, -0079909, and -0303458, the Sloan Foundation, and the Research Corporation.

Figure Captions

FIG. 1. (a) Magnetoresistance of $Pr_{0.67}Ca_{0.33}MnO_3$ at T = 35 K for both increasing (solid line) and decreasing (dotted line) field sweeps. Field pairing for insulating and low-resistance states at |B| = 3 T is indicated by solid and dashed straight lines, respectively. Increasing (decreasing) fields are filled (empty) circles. (Inset) IM breakdown (circles) and recovery (solid squares) fields obtained by point of inflection are typical of $Pr_{0.67}Ca_{0.33}MnO_3$ [24]. The hatched area marks the hysteresis region.

FIG. 2. (a) $\Delta\widetilde{\theta}$ for insulating and low-resistance states of $Pr_{0.67}Ca_{0.33}MnO_3$. (b) Field sweep of $\Delta\widetilde{\theta}$ (1500 ps, solid). Arrows give sweep direction. Dashed line re-plots $\Delta\widetilde{\theta}$ after a 180° rotation about the origin to emphasize hysteresis. DC magnetic moment (dotted) and early time $\Delta R/R$ (100 ps, circles) also shown for positive sweep directions. All data taken at T = 35 K.

FIG 3. (a) $\Delta\widetilde{\theta}$ (open circles), $\Delta\widetilde{\eta}$ (squares), and $\Delta R/R$ (solid line) for the low-resistance state (-3 T, 35 K) of $Pr_{0.67}Ca_{0.33}MnO_3$. $\Delta\widetilde{\theta}$ and $\Delta\widetilde{\eta}$ are scaled by 0.28 and –1, respectively. (b) $\Delta\widetilde{\theta}$, $\Delta\widetilde{\eta}$, and $\Delta R/R$ for the insulating state (+3 T, 35 K). $\Delta\widetilde{\theta}$ is scaled by -0.45. In both (a) and (b), the agreement between $\Delta\widetilde{\theta}$ and $\Delta\widetilde{\eta}$ for $\Delta t > 1$ ns was verified out to 7 ns (not shown). Insets: Same as (a-b) but unscaled and on an expanded time axis. (c) Exponential fit rise times for $\Delta\widetilde{\theta}$ (open circles), $\Delta\widetilde{\eta}$ (squares), $\Delta R/R$ (triangles), and estimates of $\tau_m$ (open nabla) as described in the text.

FIG. 4. (a) $\alpha = \Delta\tilde{\eta} + 0.45\Delta\tilde{\theta}$ (triangles) for the insulating state (+3 T, 35 K) in PCMO, and $\Sigma$ (open circles), obtained from inset as defined in the text. Solid line shows exponential fit to $\Sigma$. (Inset) $\Delta R/R$ vs. probe energy at $\Delta t = 40$ ps (solid line) and 4000 ps (dashed line) at B = 0 T. DC reflectivity changes $(R(T)-R(35\ K))/R(35\ K)$ $(\times 10^{-3})$ are also shown at B = 0 T for T = 80 K (closed circles) and 110 K (open circles). (b) $\alpha = (\Delta\tilde{\theta} - 0.95\Delta\tilde{\eta})$ in LCMO at T = 270 K, B = 2 T (closed circles). (c) DC Kerr effect in PCMO at T = 35 K with IM transitions marked by dashed straight lines. DC Kerr data are taken using a low-intensity incoherent CW light source at 1.55 eV and field-pair subtracted to represent a positive field sweep direction.

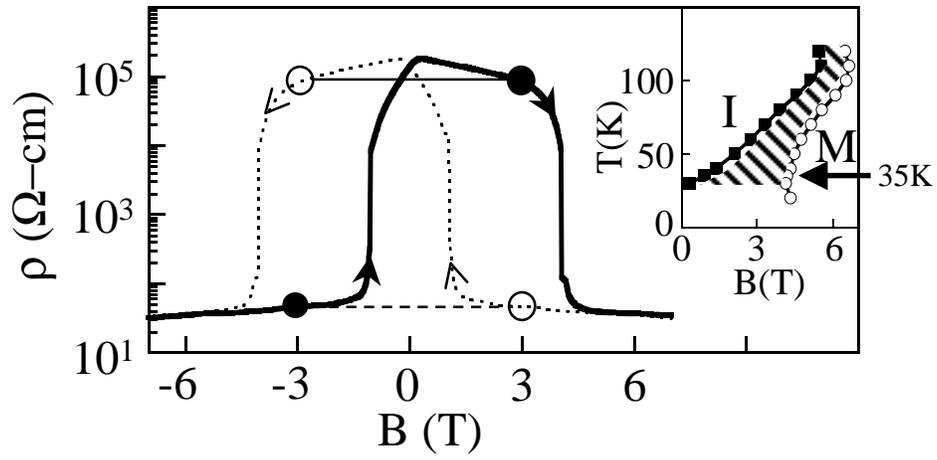

Figure 1 (McGill, *et al*)

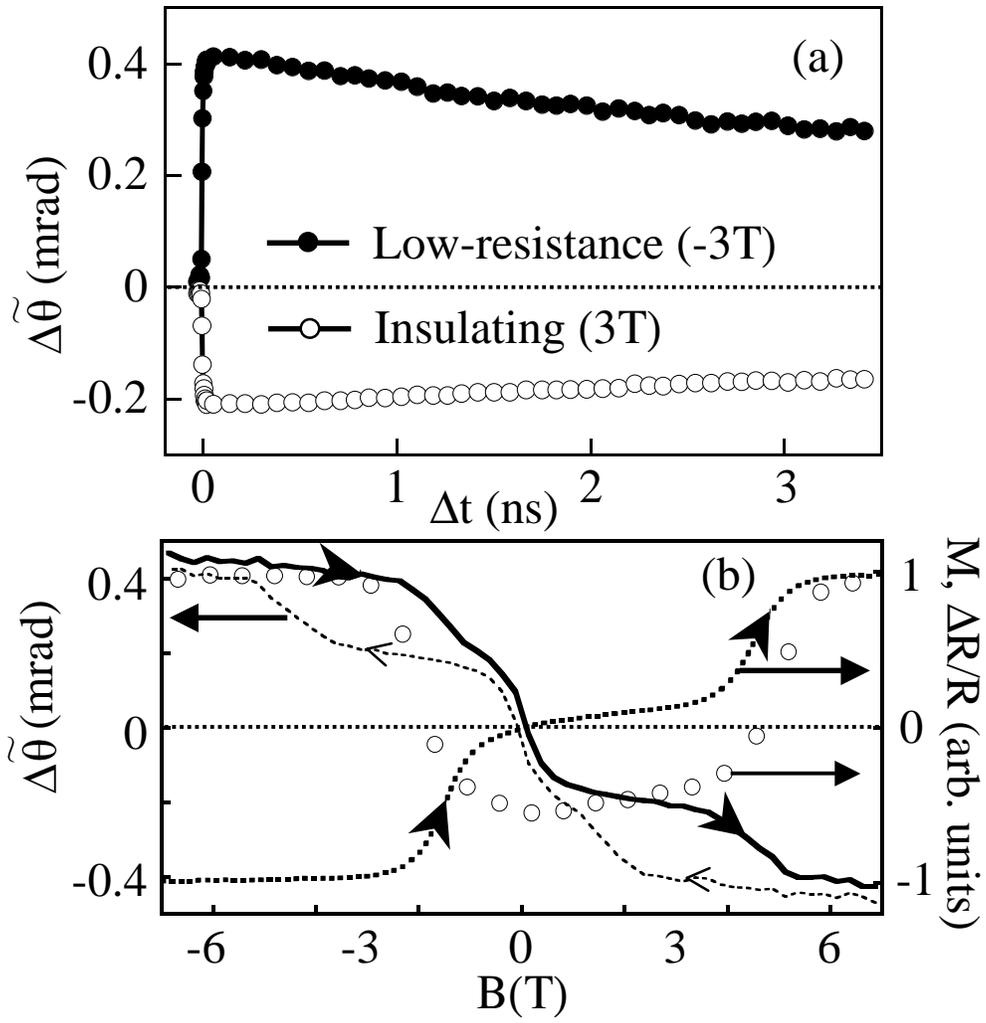

Figure 2 (McGill, *et al*)

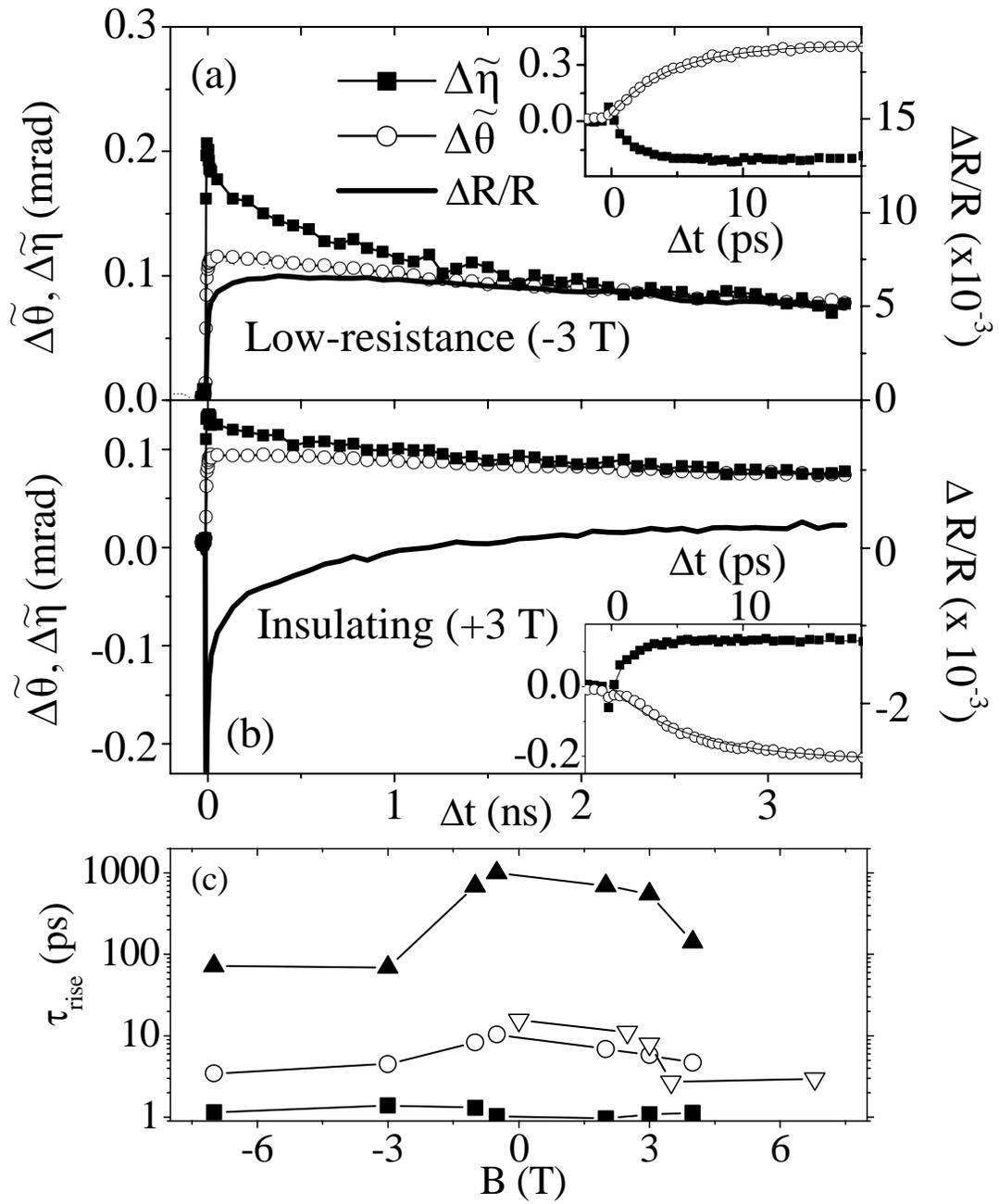

Figure 3 (McGill, *et al*)

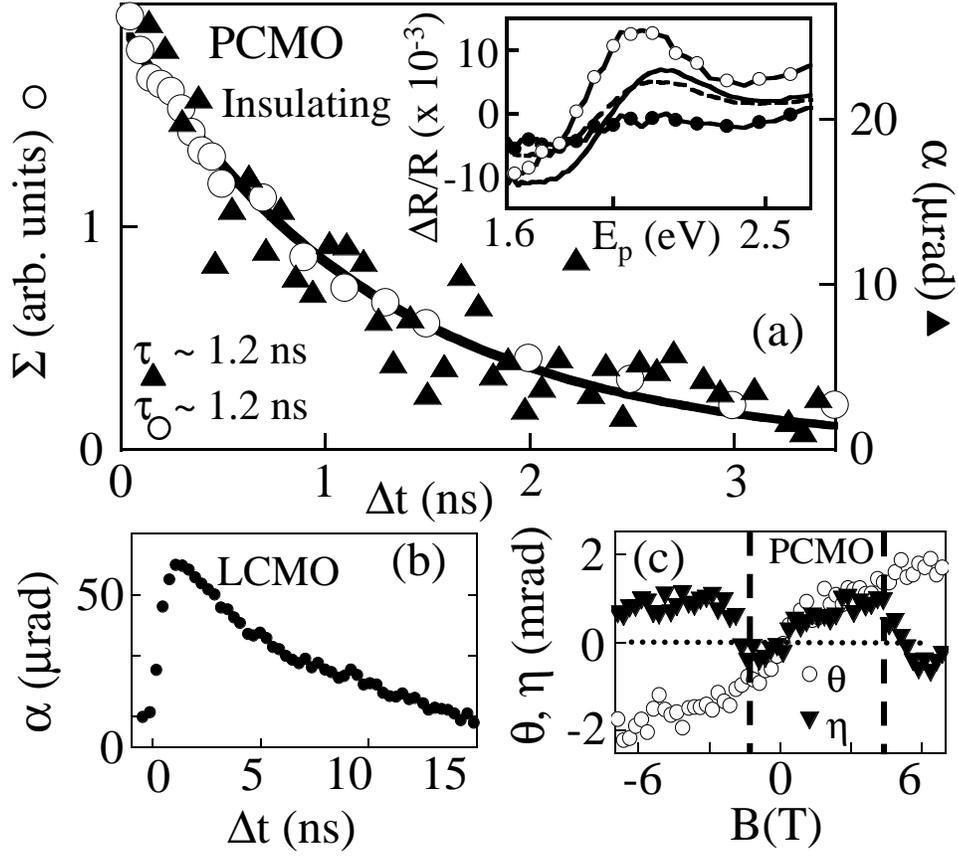

Figure 4 (McGill, *et al*)